\documentclass[10pt,journal,final]{IEEEtran}
\IEEEoverridecommandlockouts

\usepackage{amsfonts,amssymb}
\usepackage{ifpdf}
\usepackage{cite}
\usepackage{makecell}
\usepackage{multirow}
\usepackage[hyphens]{url}
\usepackage{stfloats}
\usepackage{bm}
\usepackage{color,xcolor}
\usepackage{subfigure}
\ifCLASSINFOpdf
   \usepackage[pdftex]{graphicx}
   \graphicspath{{../pdf/}{../jpeg/}}
   \DeclareGraphicsExtensions{.pdf,.jpeg,.png}
\else
   \usepackage[dvips]{graphicx}

   \graphicspath{{../eps/}}

   \DeclareGraphicsExtensions{.eps}
\fi

\usepackage[cmex10]{amsmath}
\usepackage{algorithm}
\usepackage{algorithmic}
\ifCLASSOPTIONcompsoc
\else
  \usepackage[caption=false,font=footnotesize]{subfig}
\fi
\usepackage{makecell}
\begin{document}

\title{Wireless Physical Neural Networks (WPNNs):\\ Opportunities and Challenges}
\author{Meng Hua, Itsik Bergel, Tolga Girici, Marco Di Renzo, and Deniz~G\"und\"uz

 \thanks{Meng Hua and Deniz~G\"und\"uz are with  Imperial College London, United Kingdom. (e-mail: m.hua@imperial.ac.uk); Itsik Bergel is with  Bar-Ilan University, Israel; 
Tolga Girici is with  TOBB University of Economics and Technology, Türkiye; 
Marco Di Renzo is with Universit\'e Paris-Saclay, France, and also with King's College London, United Kingdom.}
}
\vspace{-0.5cm}

\maketitle

\begin{abstract}
Wireless communication systems exhibit structural and functional similarities to neural networks: signals propagate through cascaded elements, interact with the environment, and undergo transformations. Building upon this perspective, we introduce a unified paradigm, termed \textit{wireless physical neural networks (WPNNs)}, in which components of a wireless network, such as transceivers, relays, backscatter, and intelligent surfaces, are interpreted as computational layers within a learning architecture. By treating the wireless propagation environment and network elements as differentiable operators, new opportunities arise for joint communication-computation designs, where system optimization can be achieved through learning-based methods applied directly to the physical network. This approach may operate independently of, or in conjunction with, conventional digital neural layers, enabling hybrid communication learning pipelines. In the article, we outline representative architectures that embody this viewpoint and discuss the algorithmic and training considerations required to leverage the wireless medium as a computational resource. Through numerical examples, we highlight the potential performance gains in processing, adaptability, efficiency, and end-to-end optimization, demonstrating the promise of reconfiguring wireless systems as learning networks in next-generation communication frameworks.
\end{abstract}

\section{Introduction}
Wireless communication systems inherently perform distributed, layered, and task-dependent signal transformations through electromagnetic wave propagation, hardware processing, and network interactions. As signals traverse cascaded wireless components, such as transceivers, relays, backscatter, and intelligent surfaces, they naturally undergo linear mixing, aggregation, and nonlinear distortion, closely mirroring the fundamental operations of deep neural networks (DNNs). In particular, multi-hop propagation introduces depth, antenna arrays and spatial superposition provide high-dimensional linear transformations, and radio-frequency (RF) hardware nonidealities supply intrinsic nonlinear responses. These properties suggest that wireless networks are not merely data-delivery infrastructures, but can be viewed as physical substrates that natively implement neural-like computations during signal transmission. From this perspective, conventional approaches that rely on executing DNNs purely in the digital domain treat wireless systems as passive bit pipes, overlooking their inherent computational capability. 
Biological systems operate very differently: in the brain, communication and computation are inseparable. Dendrites, synapses, and axons do not just relay signals, they actively filter, transform, and integrate them in real time. The very pathways that transmit information also compute with it, erasing any clear boundary between ``communication'' and ``processing''. This integrated design grants the brain remarkable efficiency. By compressing and reweighting data at every transmission stage, it performs complex computations without wastefully shuttling raw data between distinct modules. As AI becomes increasingly central to communication systems, maintaining this strict separation of roles may no longer be sustainable.

\subsection{What are WPNNs?}
Motivated by the observation that wireless propagation already embodies the core primitives of neural computation, we present a new paradigm, referred to as \textit{wireless physical neural networks (WPNNs)}  \cite{reus2023airfc, Garcia2023irNN, bian2025overtheair}.
WPNNs are learning architectures in which wireless propagation channels,  RF hardware, and network elements collectively define the computational graph of a DNN. In contrast to conventional digital DNNs, where computation is explicitly carried out through arithmetic operations, WPNNs embed neural parameters directly into the physical states of wireless systems, such as amplification factors in relay networks.  In WPNNs, fundamental neural operations naturally emerge from the physics of communication systems. Specifically, linear transformations arise from the superposition and mixing of signals across multiple antennas, propagation paths, and frequency resources, which are governed by the wireless channel, analog precoders/combiners, and programmable propagation elements. Network depth is introduced through cascaded signal propagation across multiple stages, such as multi-hop relays or stacked intelligent metasurfaces, where each hop or layer performs a successive transformation analogous to a neural layer. Moreover, nonlinear activation functions are inherently provided by RF hardware components, including power amplifiers (PAs), rectifiers,  transistor-based circuits, etc.,  whose saturation and thresholding behaviors emulate commonly used neural activations.  
\subsection{Key Advantages for WPNN Implementation}
The significance of WPNNs lies not in the mere realization of neural operations through physical processes, but in the distinct system-level advantages enabled by wireless propagation and networked signal transmission. These advantages fundamentally differentiate WPNNs from both digital DNNs and non-wireless physical neural computing platforms.
\begin{itemize}
  	\item \textbf{Unified computation–communication architecture:} In WPNNs, neural computation is intrinsically embedded into the wireless communication process itself. Communication primitives such as signal propagation, superposition, beamforming, and relaying simultaneously serve as computational operations, eliminating the conventional boundary between communication and computation. This convergence has already proven effective in the context of distributed learning: over-the-air federated learning exploits waveform superposition to aggregate local model updates directly during transmission, thereby turning the multiple-access channel into a natural gradient summation operator \cite{Amiri:TSP:20}. WPNNs extend this principle from training to inference, embedding neural computation itself, rather than merely parameter aggregation, into the propagation process. The resulting architecture enables joint optimization of transmission and learning objectives at the physical layer, so that communication resources directly contribute to task-level intelligence, rather than merely delivering data to external processors.
	\item  \textbf{High energy efficiency:} 
    By embedding neural parameters directly into the physical layer, WPNNs avoid explicit memory access and digital multiply-and-accumulate operations that dominate energy consumption in conventional architectures. This advantage is especially pronounced in distributed wireless systems, where communication energy dominates overall system power.
	\item \textbf{Ultra-low latency:}  Neural processing in WPNNs is executed concurrently with signal transmission, rather than as a separate post-communication step, so that inference and feature extraction can be completed during propagation. This over-the-air computation capability is particularly advantageous for delay-sensitive applications such as real-time sensing, control, and edge intelligence.
    \item \textbf{Native parallelism and scalability:} Wireless systems naturally support simultaneous signal transmission from multiple nodes over a shared medium. WPNNs exploit this property to perform parallel neural computation across spatially distributed devices without centralized coordination, with computational capacity scaling organically with the network size.

\end{itemize}

\subsection{How to Implement WPNNs}
\begin{figure*}[!t]
	\centerline{ \includegraphics[width=6in]{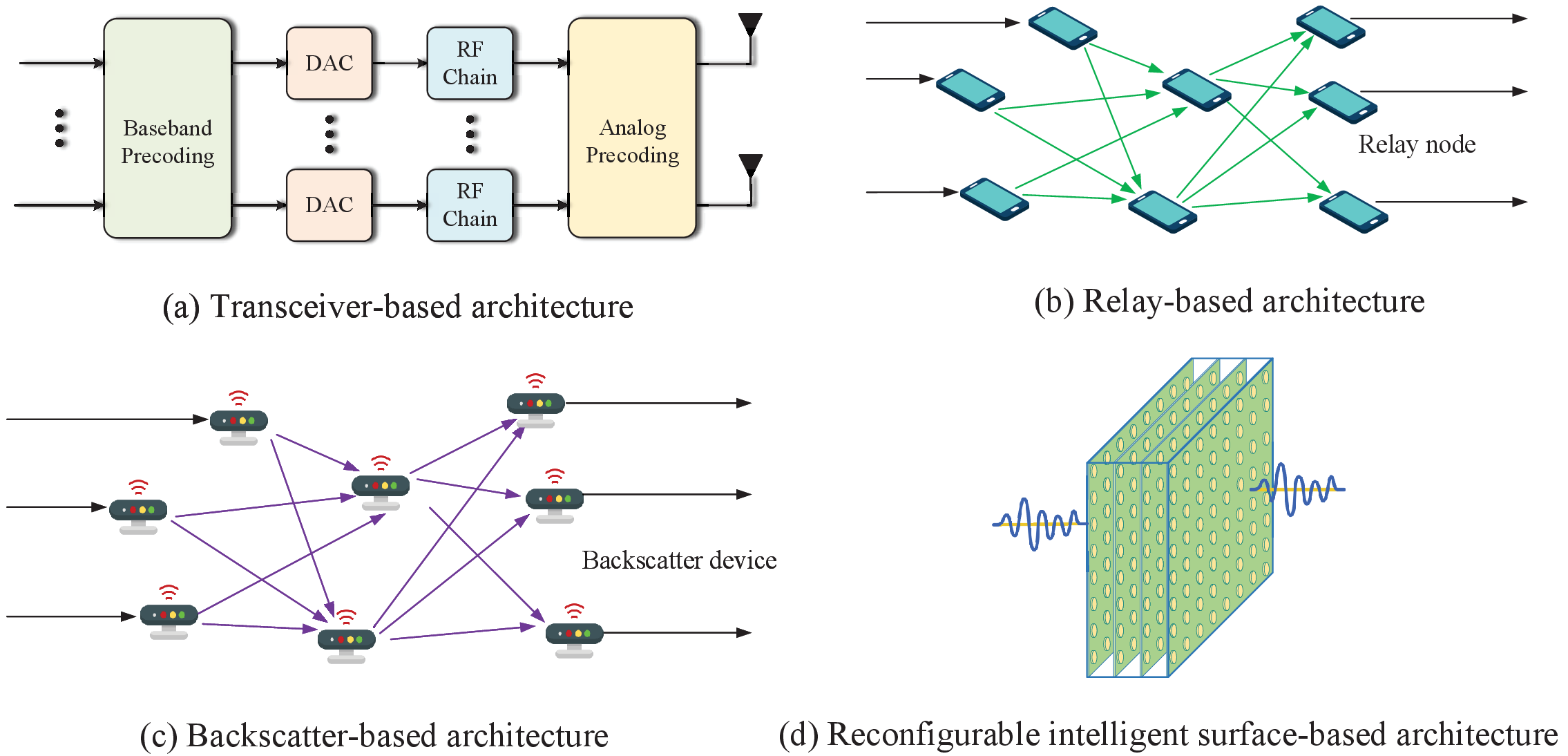}}
	\caption{\color{black}Promising wireless architectures for WPNN implementation. Transceiver-based designs exploit MIMO channel mixing for analog linear operations; Relay-assisted systems introduce active signal transformation and multi-hop depth;  Backscatter-based architectures utilize passive signal reflection and modulation to enable ultra-low-power analog computing;
    RIS-enabled approaches leverage programmable reflections for scalable energy-efficient computation.} \label{PNN:architecture}
    \vspace{-0.4cm}
\end{figure*}

Implementing WPNNs relies on mapping neural computations, such as weighted summations and matrix multiplication, onto the intrinsic propagation and interaction mechanisms of the wireless medium. The wireless channel, rather than serving merely as a communication medium, becomes part of the computing substrate. Depending on the network architecture and deployment scenario, WPNNs can be physically realized through four representative wireless structures: transceiver 
, relay, backscatter, and reconfigurable intelligent surface (RIS), as shown in Fig.~\ref{PNN:architecture}. 
Some initial research on  WPNNs has been conducted. For example, the RIS-based hardware for implementing WPNNs has been studied in \cite{stylianopoulos2025over,hua2025aircnn,hua2025implementing}, where the phase shift itself is treated as part of the NN parameters, which can be directly trained. 
In addition, the multi-hop relay-based hardware for implementing WPNNs has been studied in \cite{bergel2024nonlinear,bian2025overtheair}, where the results show that multi-hop relay can implement simple inference tasks. 

In addition to the linear analog operations enabled by wireless propagation, the implementation of nonlinear activation functions is also critical for achieving expressive WPNNs. Without appropriate nonlinear mechanisms, the computational capability of  WPNNs would remain fundamentally limited, regardless of the advantages offered by the wireless medium. To this end, the nonlinear elements, such as diode-based rectifiers and PAs, provide intrinsic thresholding or saturating responses. For example,  work \cite{bergel2024nonlinear} showed that the non-linear behavior of a PA is very similar to the hyperbolic tangent (tanh) activation function. These devices offer a highly attractive pathway for scalable and practical WPNN activation, complementing the inherent linearity of over-the-air computation.


Despite the promising initial efforts summarized above, the design of WPNNs still remains in its early stages, and several fundamental questions are still open: which wireless architectures are best suited to realize specific neural operations, how expressive the resulting physical networks can be, and what training and deployment challenges must be addressed before WPNNs can be considered practical. This article aims to provide a unified treatment of these questions by interpreting the core components of a wireless network, such as transceivers, relays, backscatter devices, and RISs, as configurable computational layers whose tunable physical parameters serve as learnable neural weights. Within this framework, the wireless propagation environment and network elements are treated as differentiable operators, opening new opportunities for joint communication-computation designs, in which system optimization is achieved through learning-based methods applied directly to the physical network.

\section{Architecture Design for WPNNs}
In this section, we present four representative hardware architectures for implementing WPNNs, namely transceiver-based, relay-assisted, backscatter-enabled, and RIS-enabled designs. For each architecture, we describe the underlying operating principles, associated signal models, and communication-system design perspectives.
\subsection{Transceiver-based WPNN Implementation}
Transceiver-based WPNNs treat the cascade of the analog front-end and wireless channel as a learnable linear transformation \cite{yang2023over}. By jointly configuring the analog precoder at the transmitter and the analog combiner at the receiver, matrix multiplications can be performed directly over the air, with the propagation medium acting as part of the computation. Under this interpretation, a fully-connected (FC) layer can be emulated in the analog domain, where the effective weight matrix is jointly determined by the precoder, channel, and combiner, while bias terms can be introduced via fixed DC offsets after combining. Practical implementations typically rely on phase shifters and variable-gain amplifiers, subject to hardware constraints such as constant-modulus or finite-resolution control.

Beyond FC layers, transceiver-based WPNNs can also emulate convolutional neural networks  (CNNs)  by exploiting the frequency and spatial degrees of freedom (DoF) of wireless systems \cite{hua2025aircnn}. In multicarrier transmissions such as orthogonal frequency-division multiplexing (OFDM), convolution-like operations can be realized in the frequency domain, where different subcarriers collectively form an effective convolutional filter. From this perspective, a MIMO-OFDM transceiver naturally functions as a multi-kernel convolutional layer, with the channel responses and analog precoding jointly shaping the filter behavior. Multiple output feature maps can be generated by partitioning the antenna array into subarrays and applying spatial beamforming.

The fidelity of transceiver-based WPNNs is ultimately constrained by the wireless channel, as fading and path loss distort the effective weight matrix. To enrich the available DoF, emerging flexible antenna technologies enable the physical antenna geometry itself to be optimized, thereby actively reshaping the channel. Movable or fluid antennas primarily affect small-scale fading, while pinching antennas modify large-scale propagation characteristics by altering the dominant scattering geometry. Joint optimization of precoding, combining, and antenna configuration thus provides an effective means to enhance the expressiveness and performance of transceiver-based WPNNs.

\subsection{Relay-based WPNN Implementation}
Relays have traditionally been used to extend coverage and improve link quality in wireless communications. In WPNNs, relays can instead be interpreted as intermediate computing layers that participate in distributed neural processing \cite{bian2025overtheair}. Heterogeneous terminals, such as repeaters, unmanned aerial vehicles (UAVs), and access points, can act as distributed relays to form multi-layer WPNNs, where each relay stage corresponds to one neural layer. In particular, amplify-and-forward (AF) relays can be viewed as tunable analog neurons, whose amplification gain matrix emulates the linear weights of an FC layer. Cascading multiple relay stages results in an equivalent virtual MIMO transformation, enabling multi-layer neural mappings to be realized directly over the air.

Relay-based WPNNs can further emulate multi-layer CNNs by exploiting multi-hop and multicarrier transmission. In OFDM systems, subcarriers can be interpreted as elementary convolution kernels, while the joint response across selected subcarriers forms an effective filter determined by cascaded relay coefficients. Stacking relay stages across hops enables successive convolutional layers, and multiple output feature channels can be generated by partitioning the relay ensemble into groups.

In addition, relay mobility provides an extra physical DoF to reshape the propagation environment. For example, UAV-based relays can adjust their three-dimensional positions to establish favorable links and better match the desired analog transformation of each neural layer, thereby enhancing the performance of multi-layer WPNNs.

\subsection{Backscatter-based WPNN Implementation}
A backscatter device typically consists of a simple antenna and a tunable impedance network. By switching its reflection coefficient or modulation pattern, the tag perturbs the incoming waveform and, in effect, reshapes the channel response between a transmitter and a receiver. This controllable reflection turns the backscatter link into a lightweight analog processing node rather than a mere passive reflector.

In a backscatter-based WPNN, each tag can be viewed as a small analog neuron whose reflection coefficient or modulation state implements a learnable weight. When an illuminating signal propagates through a field of tags, the superposition of their reflected components at the receiver realizes a weighted sum of the incident waveform, analogous to a digital FC layer. By coordinating multiple layers or groups of backscatter devices, each with different reflection settings, the system can form multi-layer FC mappings over the air, while consuming orders-of-magnitude less power than active relays. 

The extension to CNNs follows a similar principle to the AF-relay case, where different frequency tones or spatial groups of tags can be associated with distinct convolution kernels and filters. Since this design largely behaves like the relay-based CNN realization, we do not elaborate further and instead highlight backscatter as a promising, ultra-low-power substrate for large-scale, dense WPNN deployments.

\subsection{RIS-based WPNN Implementation}
An RIS is a programmable metasurface whose elements can independently adjust their phase shift, amplitude, and reflection–transmission ratio, enabling dynamic control of the wireless propagation environment. While a single RIS panel reshapes the channel between a transmitter–receiver pair, stacking multiple panels into a stacked intelligent metasurface (SIM) \cite{10279173} creates a cascade of learnable analog transformations in the wave domain. This structure maps naturally onto a multi-layer WPNN. In this interpretation, each metasurface layer applies a diagonal phase-shift matrix interleaved with free-space propagation, and the composite effect of the full stack constitutes a virtual MIMO transformation that serves as the analog counterpart of a deep digital weight matrix.  The SIM thus performs multi-layer inference during signal propagation itself, without any explicit digital computation.

This architecture supports several neural structures. For FC-type processing, the cascaded RIS channel response implements a dense linear transformation across antenna elements, with the per-element phase shifts playing the role of trainable weights. Convolutional processing can be realized by exploiting the frequency-domain structure of OFDM: configuring the RIS phase profiles independently across subcarriers yields frequency-selective analog filters that emulate convolutional kernels, while parallel SIM stacks can be deployed to provide multiple output feature channels.

Two extensions significantly broaden the design space. First, \emph{active RIS} augments each element with a low-cost negative-resistance component, enabling simultaneous phase and amplitude control. This extension allows the metasurface to compensate for propagation losses between layers in deep cascaded architectures. Second, the physical placement of RIS panels offers a spatial DoF that has no direct counterpart in digital DNNs: by optimizing panel positions jointly with their phase configurations, the effective channel matrices between layers can be shaped to improve the conditioning and rank of the composite weight matrix, expanding the set of neural transformations realizable over the air.

\begin{table*}[t]
\centering
\caption{ Comparison of Wireless Hardware Architectures for WPNN Implementation}\label{architecture_comp}
\begin{tabular}{c|c|c|c|c|c|c|c|c}
\hline
\textbf{Architecture} & \makecell{\textbf{Native} \\ \textbf{Operation}} & \makecell{\textbf{CSI Acquisition}\\ \textbf{Difficulty}} & \makecell{\textbf{Controllable} \\\textbf{Parameters}} & \makecell{\textbf{Network}\\ \textbf{Depth}} & \textbf{Latency }  & \textbf{Energy}  & \makecell{\textbf{Hardware} \\ \textbf{Complexity}} & \makecell{\textbf{Activation }\\ \textbf{Capability}} \\ 
\hline
Transceiver      & FC, CNN & Low      & Precoder, Combiner & Low & Medium & Medium& Medium & Yes \\ 
AF Relay    & FC, CNN & Medium & Gain               & High & High & High  & High  &  Yes\\ 
Backscatter    & FC, CNN & Medium & Reflection, Modulation               & Low & Low &Low & Low    &  Limited\\
RIS          & FC, CNN & High     & Phase shift        & Medium & Low &Low  & Medium  & Limited   \\ 
\hline
\end{tabular}
\end{table*}
\subsection{Overall architecture implementation comparison}
Table~\ref{architecture_comp} compares four wireless hardware architectures for WPNN implementation from several key perspectives. While all architectures can natively support FC- and CNN-type operations, they exhibit distinct trade-offs in terms of depth, channel state information (CSI) requirements, latency, energy consumption, hardware complexity, and activation capability. Transceiver-based designs require relatively low CSI accuracy and mainly tune precoders and combiners, offering shallow network depth with moderate latency and complexity, but strong activation enabled by RF nonlinearities such as PAs. AF relay–based architectures can support very deep WPNNs by introducing multi-hop depth and, at the cost of higher CSI demand, latency, energy consumption, and hardware complexity, while also providing rich nonlinear activation through amplification circuits. Backscatter-based architectures rely on passive reflection, resulting in ultra-low energy consumption and hardware complexity with low latency, but limited activation capability. RIS-based architectures reshape propagation primarily via phase control, achieving moderate effective depth with low latency and power consumption, but requiring high-dimensional CSI and exhibiting limited activation capability. Overall, transceivers and AF relays favor expressive and deep analog computation, whereas backscatter and RIS architectures are better suited for large-scale deployments, highlighting the potential of hybrid WPNN architectures in practice.

\section{Nonlinear Hardware Primitives for WPNNs}
While Section II introduces the feasible hardware architectures for realizing WPNNs over the air, those implementations primarily focus on linear signal transformations. However, linear operations alone are not sufficient for universal function approximation, a crucial property of DNNs. In conventional digital DNNs, this capability arises from nonlinear activation functions, such as ReLU and tanh. To achieve comparable expressive power in WPNNs, it is therefore necessary to incorporate nonlinear behavior directly into the WPNN implemented on hardware. This section explores the realization of nonlinear functions through diverse physical devices and mechanisms, including the PA, rectifier, and bipolar junction transistors. These nonlinear hardware primitives serve as the analog counterparts of digital activation layers, allowing WPNNs to perform nonlinear mappings directly in the analog domain, thereby bridging the gap between physical computation and neural intelligence.

\subsubsection{Tanh implementation}
\begin{figure}[!t]
	\centerline{ \includegraphics[width=3.6in]{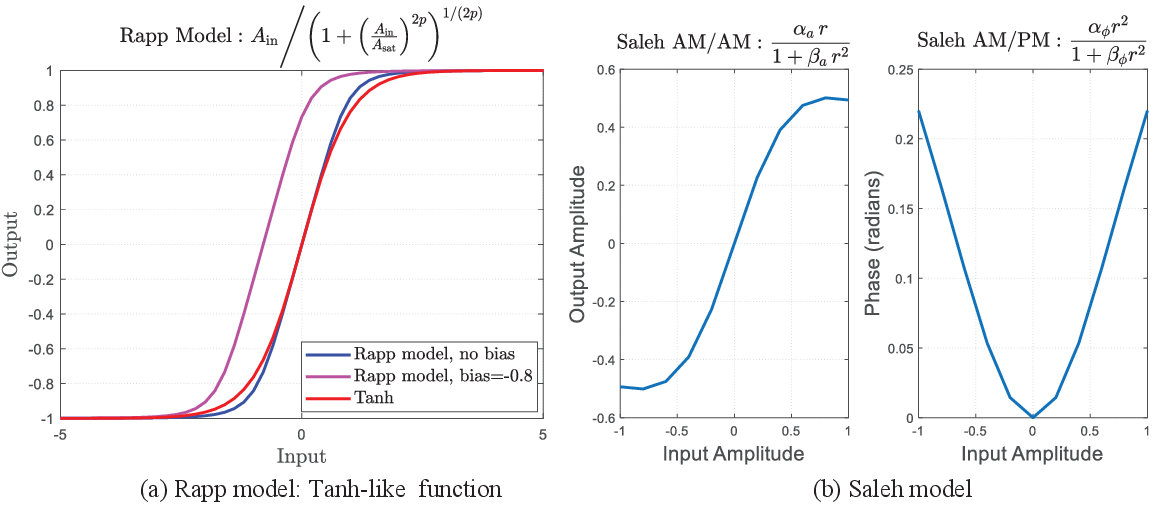}}
	\caption{Two representative PA models and their nonlinear characteristics. 
Subfigure (a) shows the Rapp model exhibiting typical activation behavior: 
a tanh-like function with $A_{\mathrm{sat}}=1$ and $p=2$. 
Subfigure (b) illustrates the Saleh model with ${{\alpha _a=1.2}}$, ${{\beta _a}=1.43}$,  ${{\alpha _\phi }=0.37}$, and ${{\beta _\phi }=0.68}$.    } \label{non_linear_activation}
    \vspace{-0.4cm}
\end{figure}
The PA plays a crucial role in wireless transceivers, providing the necessary signal power for transmission. However, practical PAs exhibit nonlinear characteristics due to device saturation, where the output amplitude gradually flattens as the input increases. The nonlinearity of PAs significantly affects the realization of WPNNs, where such nonlinear responses can be utilized as hardware activation functions. To accurately characterize these effects, several analytical models have been proposed, among which the Rapp and Saleh models are the most representative \cite{joung2014survey}. 

Fig.~\ref{non_linear_activation} illustrates representative PA models and their nonlinear behaviors. Fig.~\ref{non_linear_activation}(a) shows the Rapp model exhibiting tanh-like activation; Fig.~\ref{non_linear_activation}(b) presents the Saleh model with smooth amplitude/amplitude and amplitude/phase responses. These models demonstrate how PA nonlinearities can emulate common activation functions for WPNNs.


\subsubsection{ReLU implementation}
ReLU activation can be realized using rectifying and thresholding devices that suppress negative inputs while linearly amplifying positive ones \cite{huang2020analog,ning2025multilayer}. Diode-based rectifiers perform half-wave rectification by conducting only for positive voltages, whereas diode-connected MOSFETs achieve a similar effect through their threshold-controlled conduction, closely matching the ReLU behavior. In addition, envelope detectors also provide analog approximations by clipping negative signals while preserving linearity for positive inputs.  By adjusting biasing conditions, they can also implement leaky-ReLU variants for flexible activation in WPNNs. Moreover, memristors offer a compact and tunable approach to realizing ReLU functions, leveraging their intrinsic threshold-switching current-voltage characteristics to suppress sub-threshold signals while enabling near-linear conduction above the programmed threshold, which can also be adapted for leaky-ReLU behavior.

\begin{table*}[t]
\centering
\caption{ Nonlinear Hardware Primitives and Their Activation Capabilities in WPNNs}\label{nonlinear_activation_comp}
\begin{tabular}{c|c|c|c|c}
\hline
\textbf{Hardware Primitive} &
\textbf{Supported Activation-like Functions} &
\textbf{Power Consumption} &
\textbf{Hardware Cost} &
\textbf{Hardware Complexity} \\ 
\hline

Power Amplifier &
Tanh, Sigmoid &
High &
Medium &
Medium--High \\
\hline

Diode-based  MOSFET Rectifier &
ReLU, Leaky-ReLU &
Very Low &
Very Low &
Low \\
\hline

Envelope Detector &
 Clipped activation &
 Low &
 Low &
Low–Medium \\
\hline

Memristor&
ReLU, Sigmoid, Tanh Components&
very Low &
Low &
Medium\\
\hline
\end{tabular}
\end{table*}


Table~\ref{nonlinear_activation_comp} summarizes key nonlinear hardware primitives for implementing activation-like functions in WPNNs. The PA supports the tanh function but exhibits high power consumption. In contrast, the diode-based MOSFET rectifier efficiently realizes ReLU and Leaky-ReLU with very low power and cost. The envelope detector provides clipped activation at low power.  The memristor offers versatile ReLU, sigmoid, and tanh components at very low power. These primitives demonstrate inherent trade-offs among power, hardware cost, and complexity, guiding the selection of appropriate components for efficient WPNN implementations.

\section{Main Design Challenges for implementing WPNNs}
In this section, we will discuss main design approaches and challenges for WPNN implementation, including WPNN training, emulation, CSI acquisition, and noise accumulation.

\subsection{WPNN Training}\label{ss:WPNN_training}
Training WPNNs fundamentally differs from training conventional digital DNNs because the computations are governed by analog physical processes rather than idealized digital operations. In digital DNNs, parameters are updated through backpropagation using precise gradients computed in software, assuming deterministic arithmetic. In contrast, WPNNs operate within real physical systems with varying channels and even unknown hardware electromagnetic properties. There are two types of training paradigms for WPNNs \cite{momeni2025training}: physics-aware training (PAT) and in-situ training (IST). PAT relies on digital simulation that incorporates physical constraints and hardware characteristics into the training loop. Specifically, it uses a high-fidelity digital twin of the physical system to optimize the weights while accounting for all physical constraints such as transmit power, channel impairment, and RIS phase shift. After convergence, the optimized parameters are deployed onto the actual hardware. On the other hand, IST performs learning directly on the physical platform itself, using real measurements rather than simulated gradients. This approach can capture the true behavior of the system and adapt to time-varying hardware imperfections. 

However, training WPNNs still faces significant challenges. For PAT, accurately modeling the physical system behavior is extremely difficult because fabrication imperfections and environmental noise inevitably introduce discrepancies between the simulated model and the actual hardware. This \textit{simulation–reality gap} limits the generalization of the trained model and degrades inference accuracy when deployed on real devices. For IST, obtaining gradients within analog systems is inherently challenging, as exact derivatives cannot be directly measured and must instead be estimated through perturbation or feedback-based methods. Moreover, hardware-related issues such as device variability, calibration drift, and limited controllability of analog states further compromise reproducibility and scalability. Consequently, achieving a robust trade-off among accuracy, stability, and training efficiency, while accommodating the time-varying nature of physical systems, remains one of the key obstacles for realizing practical and large-scale WPNN training.

\subsection{WPNN Emulation}
WPNN emulation \cite{bian2025overtheair} offers a training-free implementation paradigm, where the forward inference of a pretrained digital DNN is directly realized over the wireless medium by exploiting the channel and reconfigurable hardware as physical transformations. However, emulation-based WPNNs face several intrinsic challenges. First, pretrained models rely on well-conditioned numerical representations, whereas physical signals are subject to strict power and dynamic-range constraints, requiring careful feature scaling and normalization to preserve inference fidelity. Second, digital nonlinear operations, such as normalization, cannot be naturally supported by over-the-air superposition and hardware properties.  Third, emulating high-dimensional layers often necessitates operator decomposition across multiple transmissions or hardware resources, which limits scalability and increases implementation overhead.

\subsection{CSI Acquisition}
Performance of WPNNs depends critically on accurate CSI, which plays a crucial role in both training and inference \cite{binyamini2025estimating}. During PAT, the digital twin must faithfully represent the real propagation environment, while during inference, the effective weight matrix of each physical layer is determined by the instantaneous channel realization, so that CSI errors translate directly into computational errors throughout the network.

Accurate CSI acquisition is already challenging in conventional wireless systems, but WPNNs introduce additional difficulties along several dimensions. First, many of the network elements that constitute WPNN layers, such as backscatter devices and RIS panels, are passive or semi-passive, and lack baseband processing, so their associated channels  must be estimated indirectly through external pilot signaling or feedback from other nodes, incurring additional latency, substantial overhead, and estimation error. Second, CSI dimensionality scales with the total number of tunable elements. In deep architectures comprising hundreds of RIS elements, relay nodes, and antenna ports, the resulting parameter space can be orders of magnitude larger than in a conventional MIMO link, rendering traditional pilot-based schemes impractical.
Third, when heterogeneous components coexist within the same WPNN, mutual reflections and multi-hop signal paths create tightly coupled channel interactions that cannot be decomposed into independent per-link estimates. Addressing these challenges will require new CSI acquisition strategies that exploit the structural priors of WPNNs; for example, by estimating composite per-layer transfer matrices rather than individual element-wise channels, as well as learning-based estimation methods that can be jointly trained with the WPNN parameters themselves.

\subsection{Noise Accumulation Across Physical Layers}

A fundamental distinction between digital DNNs and WPNNs is that computation in the latter is inherently stochastic: every physical layer injects additive noise into the signal as it propagates through the network. Consider a WPNN of depth $L$, where the $l$-th layer applies a linear transformation $\mathbf{W}_l$ followed by a nonlinear activation function, with additive noise injected at each stage. Because each noise term is transformed by all subsequent layers, the effective noise at the output is a nonlinear function of the noise injected at every preceding stage. In the linear regime (i.e., when the activations are approximately linear around the operating point), the total output noise power grows roughly as $\sum_{l=1}^L \|\prod_{k=l+1}^L \mathbf{W}_k\|^2 \sigma_l^2$, where $\sigma_l^2$ is the per-layer noise variance. If the weight matrices have spectral norms greater than unity, which is desired to maintain signal strength across lossy propagation stages, the noise contribution from early layers is exponentially amplified, imposing a practical upper bound on the useful depth of the network. This noise-accumulation effect has no counterpart in digital DNNs and introduces a fundamental accuracy–depth tradeoff that is unique to physical neural architectures.

\section{Numerical Case Studies}
In this section, we present numerical examples to validate the effectiveness of FC-type and CNN-type WPNN implementation over wireless communication systems. The Rapp PA model is adopted as the activation function. The Fashion MNIST dataset is considered for image classification.

 \begin{figure}[!t]
	\centerline{\includegraphics[width=3.5in]{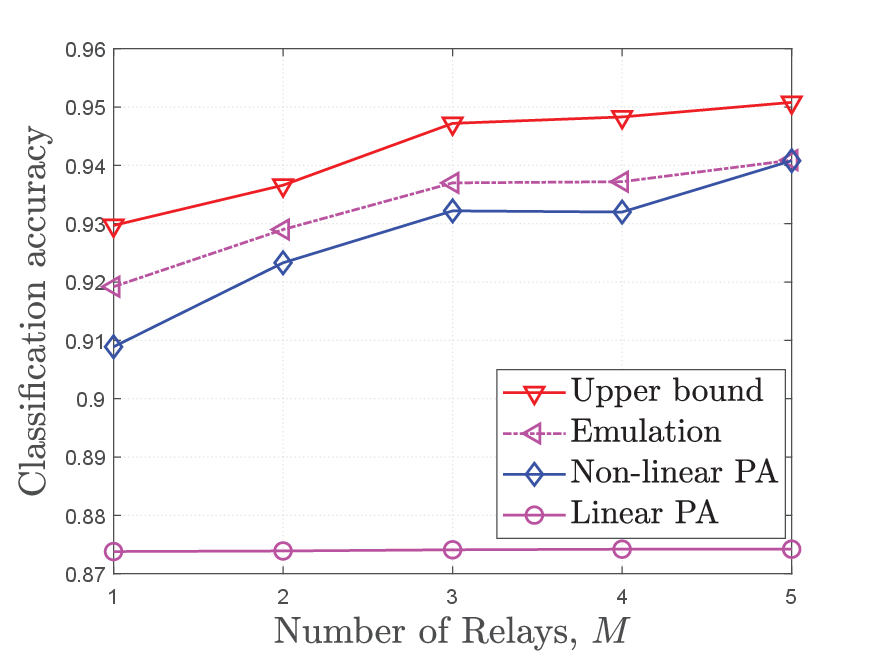}}
	\caption{Classification accuracy versus number of relays.}  \label{Relay_vs_CA}
		\vspace{-0.4cm}
\end{figure}

\subsection{Multi-FC  WPNN Implementation}

In Fig.~\ref{Relay_vs_CA}, we study the classification accuracy versus the number of relays $M$ under  ${\rm SNR}=30$~dB for different schemes. Each relay is equipped with $32$ antennas, and each relay hop can be treated as one FC layer. The ``Upper bound'' scheme corresponds to fully digital FC neural networks without channel and noise effects, whose number of FC layers is set equal to the number of relays. The ``Emulation'' scheme designs the relay amplify matrices to approximate the weight matrices obtained from the ``Upper bound'' scheme at each channel realization. The linear PA and non-linear PA schemes represent ideal linear and practical Rapp PA models, respectively. It can be observed that the accuracy of all schemes except linear PA improves with $M$. In particular, the accuracy of non-linear PA scheme increases from 0.909 at $M=1$ to 0.941 at $M=5$, approaching the upper bound. Linear PA scheme remains nearly constant since cascading multiple linear FC layers without non-linear activation cannot benefit from the depth.
We also considered a mismatched PA scheme, where the model is trained assuming a linear PA while testing is conducted with a nonlinear PA. We observed significantly lower accuracies of $0.48$ for $M=1$ and $0.30$ for $M=5$. The degradation in accuracy in this case is attributed to the accumulation of mismatch-induced nonlinear distortion across multiple hops. 

The nonlinear PA result confirms that we can exploit hardware-induced nonlinearity as an essential computational resource. It provides the activation function that can turn depth into increased expressivity. The severe degradation under PA mismatch underscores the importance of physics-aware training, as failing to account for the true hardware response during training (simulation–reality gap discussed in Section \ref{ss:WPNN_training}) leads to compounding errors across layers.

\subsection{CNN-type PNN Implementation}

\begin{figure}[!t]
	\centerline{\includegraphics[width=3.1in]{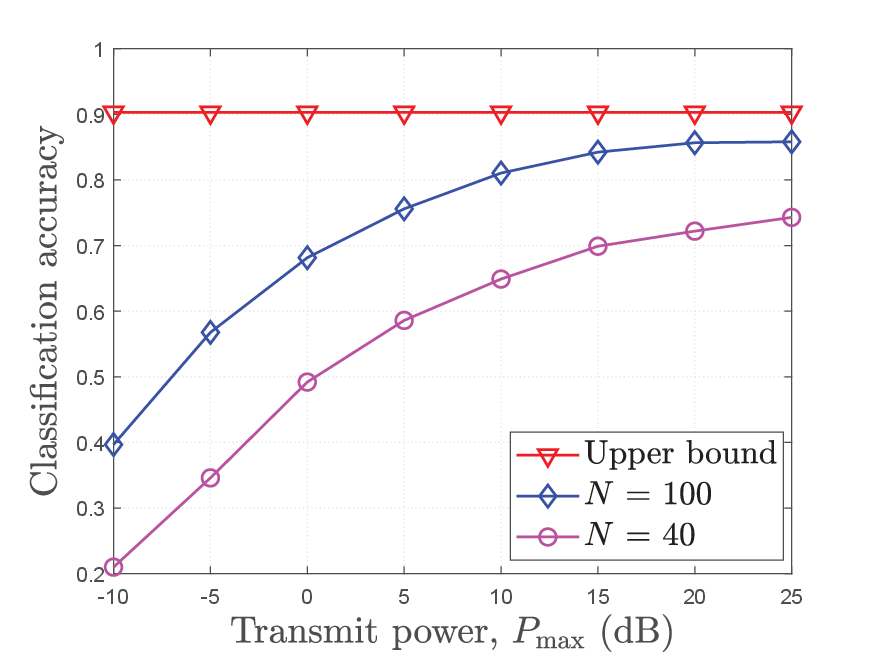}}
	\caption{Transmit power versus classification accuracy.}  \label{Simul:RISVSCA}
    \vspace{-0.4cm}
\end{figure}
In Fig.~\ref{Simul:RISVSCA}, we investigate the classification accuracy versus the transmit power $P_{\max}$ for different numbers of RIS reflecting elements $N$. The considered CNN has $32$ input and $32$ output channels with a convolutional kernel size of $3$. The transceiver is combined with a single RIS to emulate the behavior of a CNN in the digital domain. 
We adopt $32$ OFDM subcarriers, and the transmitter and receiver are equipped with $9$ and $64$ antennas, respectively.
The ``Upper bound'' scheme corresponds to the digital-domain convolution without over-the-air computation. It can be observed that the classification accuracy increases with the transmit power $P_{\max}$ for both  $N=40$ and $N=100$. A larger number of RIS elements $N$ achieves consistently better performance since increasing 
$N$ provides more spatial DoF and stronger beamforming gain. Notably, the gap to the digital upper bound narrows rapidly with transmit power, suggesting that at moderate-to-high SNR the dominant bottleneck shifts from noise to the fidelity of the analog weight approximation. In this regime, CSI acquisition becomes the performance-limiting factor.

\section{Conclusion}
This article presented a unified framework for WPNNs, in which transceivers, relays, backscatter devices, and RISs are reinterpreted as learnable computational layers whose tunable physical parameters serve as neural weights. We have shown how FC- and CNN-type architectures can be realized over the air through the interplay of analog precoding, multi-hop propagation, and programmable metasurfaces, with nonlinear activation provided by intrinsic RF hardware responses. Numerical results confirm that properly designed WPNNs can approach the performance of their digital counterparts, particularly when nonlinear activation and sufficient network depth are available. Several fundamental challenges remain open and warrant further investigation: establishing formal expressivity and approximation bounds for depth- and noise-limited physical architectures; developing scalable training algorithms that bridge the simulation–reality gap; and designing CSI acquisition methods whose overhead does not diminish the latency and energy advantages that motivate WPNNs in the first place. Addressing these questions will be essential to advancing WPNNs from a conceptual paradigm to a practical building block of next-generation intelligent wireless networks.

\bibliographystyle{IEEEtran}
\bibliography{PNN_reference}

@inproceedings{huang2020analog,
  title={Analog circuit implementation of neurons with multiply-accumulate and {ReLU} functions},
  author={Huang, Yucong and others},
  booktitle={Great Lakes Symposium on VLSI},
  pages={493--498},
  year={2020}
}

@article{hua2025implementing,
  title={Implementing Neural Networks Over-the-Air via Reconfigurable Intelligent Surfaces},
  author={Hua, Meng and Bian, Chenghong and Wu, Haotian and Gunduz, Deniz},
  year={early access, 2026},
journal = {IEEE Trans. Wireless Commun.}
}

@INPROCEEDINGS{10279173,
  author={An, Jiancheng and Di Renzo, Marco and Debbah, Mérouane and Yuen, Chau},
  booktitle={Proc. IEEE Int. Conf. Commun.  (ICC )}, 
  title={Stacked Intelligent Metasurfaces for Multiuser Beamforming in the Wave Domain}, 
  year={ Rome, Italy, 2023},
  pages={2834-2839},
}

@article{hua2025aircnn,
  title={{AirCNN} via Reconfigurable Intelligent Surfaces: Architecture Design and Implementation},
  author={Hua, Meng and Wu, Haotian and G{\"u}nd{\"u}z, Deniz},
  year={2025},
Publisher                = {[Online]. Available:},
 Url                      = {https://arxiv.org/abs/2510.25389.}
}

@article{joung2014survey,
  title={A survey on power-amplifier-centric techniques for spectrum-and energy-efficient wireless communications},
  author={Joung, Jingon and Ho, Chin Keong and Adachi, Koichi and Sun, Sumei},
  journal={IEEE Comm. Sur. Tutor.},
  volume={17},
  number={1},
  year={2014},
  publisher={IEEE}
}

@Article{momeni2025training,
  Title                    = {Training of physical neural networks},
  Author                   = {Momeni, Ali and others},
  Journal                  = {Nature},
  Year                     = {2025},
  Number                   = {8079},
  Pages                    = {53--61},
  Volume                   = {645},
}

@ARTICLE{bergel2024nonlinear,
  author={Bergel, Itsik},
  journal={IEEE Trans. Wireless Commun.}, 
  title={Non-Linear Relay Optimization Using Deep-Learning Tools}, 
  year={2024},
month={Dec.},
  volume={23},
  number={12},
}

@Article{stylianopoulos2025over,
  Title                    = {Over-the-Air Edge Inference via End-to-End Metasurfaces-Integrated Artificial Neural Networks},
  Author                   = {Stylianopoulos, Kyriakos and Di Lorenzo, Paolo and Alexandropoulos, George},
  Year                     = {2025},
  Publisher                = {[Online]},
  Url                      = {https://arxiv.org/abs/2504.00233.}
}

@Article{Garcia2023irNN,
  Title                    = {{AirNN}: Over-the-Air Computation for Neural Networks via Reconfigurable Intelligent Surfaces},
  Author                   = {Garcia Sanchez, Sara and others},
  Journal                  = {IEEE/ACM Tran. Netw.},
  Year                     = {2023},

  Month                    = {Dec.},
  Number                   = {6},
  Pages                    = {2470-2482},
  Volume                   = {31}
}

@ARTICLE{bian2025overtheair,
  author={Bian, Chenghong and Hua, Meng and Gündüz, Deniz},
  journal={IEEE Wireless Communications Letters}, 
  title={Over-the-Air Inference Through Analog Computation Over Multi-Hop {MIMO} Networks}, 
  year={2025},
  month={Nov.},
  volume={14},
  number={11},
  pages={3739-3743},
}

@article{ning2025multilayer,
  title={Multilayer nonlinear diffraction neural networks with programmable and fast ReLU activation function},
  author={Ning, Yu Ming and others},
  journal={Nature Commun.},
  volume={16},
  number={1},
  pages={10332},
  year={2025},
}

@inproceedings{binyamini2025estimating,
  title={Estimating and Optimizing of Deep Relay Networks},
  author={Binyamini, Ido and Bergel, Itsik},
  booktitle={IEEE Int'l. Works. Signal Proc. Artf. Intel. Wireless Comm. (SPAWC)},
  address={Surrey, UK},
  year={2025},
}

@Article{yang2023over,
  Title                    = {Over-the-Air Split Machine Learning in Wireless {MIMO} Networks},
  Author                   = {Yang, Yuzhi and others},
  Journal                  = {IEEE J. Sel. Areas Commun.},
  Year                     = {2023},
  Month                    = {Apr.},
  Number                   = {4},
  Pages                    = {1007-1022},
  Volume                   = {41}
}

@InProceedings{reus2023airfc,
  Title                    = {{AirFC}: Designing Fully Connected Layers for Neural Networks with Wireless Signals},
  Author                   = {Reus-Muns, Guillem and Alemdar, Kubra and Sanchez, Sara and Roy, Debashri and Chowdhury, Kaushik R},
  Booktitle                = {MobiHoc.},
  Year                     = {2023},
  Month                    = {Oct.},
  Pages                    = {71--80}
}

@ARTICLE{Amiri:TSP:20,
  author={Mohammadi Amiri, Mohammad and Gündüz, Deniz},
  journal={IEEE Trans. Signal Process.}, 
  title={Machine Learning at the Wireless Edge: Distributed Stochastic Gradient Descent Over-the-Air}, 
  year={2020},
  volume={68},
  number={},
  pages={2155-2169},
  doi={10.1109/TSP.2020.2981904}}



\end{document}